\pdfoutput=1
\documentclass[prb,twocolumn]{revtex4-1}

\usepackage{bm}
\usepackage{amssymb}
\usepackage{amsthm}
\usepackage{amsmath}
\usepackage{graphicx}
\usepackage{tabularx}
\usepackage{booktabs}
\usepackage{notes2bib}
\bibnotesetup{
note-name = ,
use-sort-key = false
}

\newcommand*{\citen}[1]{%
  \begingroup
    \romannumeral-`\x % remove space at the beginning of \setcitestyle
    \setcitestyle{numbers}%
    \cite{#1}%
  \endgroup   
}

\DeclareUnicodeCharacter{0308}{ü}
\DeclareUnicodeCharacter{0300}{è}

\begin{document}

\title{Exploring the Statically Screened $G3W2$ Correction to the $GW$ Self-Energy: Charged Excitations and Total Energies of Finite Systems}

\author{Arno F{\"o}rster}
\email{a.t.l.foerster@vu.nl}
\affiliation{Theoretical Chemistry, Vrije Universiteit, De Boelelaan 1083, NL-1081 HV, Amsterdam, The Netherlands}
\author{Lucas Visscher}
\affiliation{Theoretical Chemistry, Vrije Universiteit, De Boelelaan 1083, NL-1081 HV, Amsterdam, The Netherlands}

\date{\today}

\begin{abstract}
Electron correlation in finite and extended systems is often described in an effective single-particle framework within the $GW$ approximation. Here, we use the statically screened second-order exchange (SOX) contribution to the self-energy ($G3W2$) to calculate a perturbative correction to the $GW$ self-energy. We use this correction to calculate total correlation energies of atoms, relative energies, as well as charged excitations of a wide range of molecular systems. We show that the second-order correction improves correlation energies with respect to the RPA and also improves relative energies for many, but not all considered systems. The dynamically screened SOX term has previously been shown to consistently lower the highest occupied molecular orbital (HOMO) quasiparticle (QP) energies and to increase the lowest unoccupied molecular orbitals (LUMO) QP energies. We show here that the statically screened $G3W2$ correction consistently increases the LUMO QP energies, while no consistent trend can be observed for the HOMO levels. Also, confirming previous results, the magnitude of the correction is much smaller with the statically screened interaction, than with the dynamically screened one. qsGW by itself is shown to be an excellent method for the calculation of charged excitation of finite systems, and it can not consistently be improved upon by the $G3W2$ correction. For range-separated hybrid starting points, the description of fundamental gaps and HOMO QP energies is slightly worsened. However, tremendous improvements upon the $GW$ LUMO energies leading to almost perfect agreement with high-level coupled cluster reference methods are observed. The evaluation of the statically screened $G3W2$ correction only comes with small additional computational cost compared to $G_0W_0$ for systems with up to 100 atoms and should therefore be suitable for practical applications.
\end{abstract}

% insert suggested PACS numbers in braces on next line
\pacs{}
% insert suggested keywords - APS authors don't need to do this
\keywords{Perturbation Theory, $GW$, Quasiparticle}

\maketitle

\section{Introduction}
Electron densities, total energies and spectral functions of an interacting many-electron system can be extracted from its single-particle Green's function.\cite{Abrikosov1975, Mattuck1992, martin2016} Instead of expanding the Green's function directly in powers of the electron-electron interaction, it is usually easier to expand the irreducible self-energy and to calculate the interacting Green's function via Dyson's equation.\cite{Dyson1949} The difficulty is then to find an approximation which captures the most important correlation effects and remains computationally tractable. 

In Hedin's $GW$ approximation (GWA),\cite{Hedin1965} the self-energy is obtained as the first term of an expansion in terms of a screened electron-electron interaction,\cite{Phillips1961} where screening is (usually) calculated in the random-phase approximation (RPA).\cite{Hubbard1957} Compared to the Hartree-Fock (HF) approximation, the first-order term in the expansion of the self-energy in terms of the bare electron-electron interaction, it takes into account that the electron-electron interaction at large distances is screened by the presence of other electrons.\cite{martin2016} For properties which are mainly dominated by long-range correlation effects, such as fundamental gaps or ionization potentials in organic molecules, the GWA is therefore often very accurate.\cite{Reining2018, Golze2019} This is especially true for the most common computational approach, where the full solution of Dyson's equation is bypassed and the quasiparticle (QP) energies are calculated from a non-interacting propagator.\cite{Hybertsen1986} In the static long-range limit, the errors introduced by the neglect of self-consistency and the absence of higher order terms in the polarizability and self-energy cancel to a large extent\cite{Kotani2007} by virtue of the Ward identity.\cite{Ward1950} When short-range correlation becomes important, the GWA is less accurate.

From the $GW$ self-energy one can calculate correlation energies using the Klein functional.\cite{Klein1961} Evaluated with a Kohn-Sham (KS) Green's function, one obtains (particle-hole) RPA,\cite{Casida1995, Dahlen2006} often derived in the framework of the adiabatic connection (AC) fluctuation-dissipation theorem\cite{Ren2012a} or as a subset of terms in the coupled cluster (CC)\cite{Coester1958, Coester1960, Cizek1966, Cizek1969, Paldus1972} singles and doubles (CCD) expansion.\cite{Scuseria2008, Scuseria2013} Since the effect of charge screening dominates in this limit, the RPA describes long-range electron correlation very accurately\cite{Langreth1977}. At short electron-electron distances, charge screening is less important and the RPA drastically overestimates the magnitude of electron correlation.\cite{Singwi1968}  In this limit it is crucial to take into account higher-order contributions (vertex corrections) to the self-energy which can then dominate.\cite{Irmler2019a} These contributions are fundamentally short-ranged and become less and less important for large electron-electron distances.\cite{Jiang2007, Ren2012a}

Besides range-separation based approaches,\cite{Kurth1999, Yan2000, Angyan2005, Jiang2007,Janesko2009a, Janesko2009b, Toulouse2009, Zhu2010, Toulouse2010} the next-to-leading-order term, the second-order exchange (SOX), is therefore the most obvious starting point to treat electron correlation beyond the RPA. The most common correction is the second-order screened exchange (SOSEX) correction.\cite{DavidL.Freeman1977} SOSEX can be derived from the CC doubles (CCD) equations by neglecting ladder and "crossed-ring"\cite{Scuseria2013} diagrams.\cite{Scuseria2008, Scuseria2013} SOSEX has been applied to total energies of atoms, and molecules,\cite{Paier2012, Ren2013} solids\cite{Gruneis2009} and to the uniform electron gas\cite{Hummel2019} and several extensions to it have been proposed as well.\cite{Paier2008, Toulouse2011b,Angyan2011,Hummel2019} In another variant the SOX term is renormalized by resummation of hole-hole ladder diagrams.\cite{Engel2006, Jiang2006} Typically, these approaches improve total correlation energies\cite{Jiang2007, Gould2019} but relative energies are not necessarily improved.\cite{Ren2012a} 

The close connection between RPA correlation energies and the $GW$ method suggests, that beyond-RPA methods might also be used to improve over the $GW$ method for charged excitations. For example, Ren et al. applied SOSEX to molecules\cite{Ren2015} and Maggio and Kresse\cite{Maggio2017} used a similar method to calculate HOMO QP energies for small molecules. It is, however, more common to formulate beyond-$GW$ schemes by taking into account vertex corrections in Hedin's equations.\cite{Hedin1965} Kutepov has solved Hedin's equations in a fully self-consistent fashion including several vertex corrections.\cite{Kutepov2016} He has applied the second-order self-energy variant (coined $GW + G3W2$) to a wide range of metals, semiconductors, and insulators,\cite{Kutepov2017, Kutepov2017a, Kutepov2018, Kutepov2021, Kutepov2021c, Kutepov2022} and observed major improvements over fully self-consistent $GW$ for spectral properties using the $GW + G3W2$ approximation to the self-energy. According to his work, the vertex corrected calculations should be performed without any constraining approximations. This means, Hedin's equations should be solved fully self-consistently and the frequency dependence of all quantities should be properly accounted for. As many other authors,\cite{Onida1995, Onida2002, Adragna2003, Marini2003, Bruneval2005, Shishkin2007, Romaniello2009, Sharma2011, Chen2015, Hung2016, Cunningham2018, Olsen2019, Schmidt2017, Tal2021, Cunningham2021} he also emphasized the importance of vertex corrections in the polarizabiliy. 

On the other hand, the $GW + G3W2$ self-energy has also been applied by Gr{\"u}neis et al.\cite{Gruneis2014} to calculate QP energies in solids in a more approximate fashion. In contrast to Kutepov, Gr{\"u}neis et al.\cite{Gruneis2014} used a statically screened interaction and calculated perturbative corrections to quasiparticle self-consistent $GW$ (qsGW)\cite{VanSchilfgaarde2006, Kotani2007} QP energies. They also combined this method with a beyond RPA screening of the Coulomb interaction using a static exchange-correlation kernel. As Kutepov, they reported major improvements over the $GW$ method for the band structures of solids. Comparing their results to Kutepov's, the logical conclusion would be, that the errors introduced by the static approximation, the use of a non-interacting Green's function, and the perturbative treatment of the $G3W2$ cancel to a large extent. A recent study by Kutepov has confirmed this conjecture.\cite{Kutepov2022} From a pragmatic point of view this is of course convenient, since all of these approximations come with drastically reduced computational cost compared to the rigorous, self-consistent formalism\cite{Kutepov2016}. In a recent study, the $GW + G3W2$ self-energy has also been employed by Rinke, Ren and coworkers\cite{Wang2021} to ionization potentials and electron affinities of molecules. Rinke, Ren and coworkers used the dynamically (RPA) screened interaction and did not consider any self-consistency in solving Dyson's equation. Improvements over $GW$ were found to be substantial, especially for electron affinities and the dependence on the starting point was reduced.\cite{Wang2021}

Encouraged by these results, we herein follow Gr{\"u}neis et al.\cite{Gruneis2014} and calculate perturbative $G3W2$ corrections to the $GW$ self-energy. We follow their approach in using non-interacting Green's functions throughout and exclusively consider the statically screened $G3W2$ self-energy. However, following Rinke, Ren and coworkers\cite{Wang2021}, we do not consider any beyond-RPA screening in our work. One should expect an expansion of the polarizability in terms of $W$ to converge much faster for finite, than for extended systems, since charge screening will be much weaker.\cite{VanDenBrink2000} It is known for finite systems, that inclusion of the vertex in the polarizability alone does not improve\cite{Verdozzi1995, Schindlmayr1998} and might even deteriorate the accuracy of charged excitations compared to $GW$.\cite{Lewis2019} While Gr{\"u}neis et al.\cite{Gruneis2014} investigated periodic systems, we focus here on the calculation of correlation energies as well as QP energies for a wide range of finite systems: We show that massively improved absolute correlation energies are obtained compared to the RPA and also that relative energies are improved for the majority of the considered systems. For QP energies we show, that the $G3W2$ correction provides improvements over $G_0W_0$ evaluated with KS orbitals and energies from range-separated hybrid functionals, especially for electron affinities. However, no consistent improvements over qsGW are found which we show to be very accurate by itself.

The remainder of this work is organized as follows: In the next section, we introduce the $G3W2$ self-energy and discuss how it can be used efficiently to calculate corrections to $GW$ QP energies and RPA correlation energies. In section~\ref{sec::res}, we first benchmark the performance of the $GW + G3W2$ self-energy for absolute correlation energies as well as relative energies of several organic molecules and then proceed to scrutinize the effect of the $G3W2$ self-energy correction to $G_0W_0$ on ionization potentials, electron affinities and fundamental gaps of molecular systems. Finally, we summarize and conclude this work in section ref.~\ref{sec::conclusion}.

\section{\label{sec::theory}Theory}
Dyson's equation,\cite{Dyson1949}
\begin{equation}
    \label{DysonG}
    G(1,2) = G^{0}(1,2) + G^{0}(1,\overline{3})
    \Sigma(\overline{3},\overline{4})G(\overline{4},2) \;,
\end{equation}
relates the interacting Green's function $G$ of a many-electron system to the Green's function of a non-interacting reference system $G^0$, via the non-Hermitian self-energy operator $\Sigma$. We have used the notation $1 = (\bm{r}_1, \sigma_1, \omega_1)$ and integration over indices with upper bars is implied. Following Hubbard\cite{Hubbard1957}, Phillips\cite{Phillips1961} and Hedin,\cite{Hedin1965} the self-energy can be expanded in terms of the screened electron-electron interaction. The first two terms in this expansion are 
\begin{equation}
    \label{fullSelfEnergy}  
    \Sigma^{GW + G3W2}(1,2) = 
    \Sigma^{GW}(1,2) +
    \Sigma^{G3W2}(1,2) \;,
\end{equation}
with
\begin{equation}
    \Sigma^{GW}(1,2) =  iG(1,2)W(1,2) \;,
\end{equation}
and 
\begin{equation}
\Sigma^{G3W2}(1,2) = 
    -G(1,\overline{4})W(1,\overline{3})
    G(\overline{4},\overline{3})
    G(\overline{3},2)W(\overline{4},2) \;.
\end{equation}
$W$ is the screened Coulomb interaction which can be obtained from the polarizability $P$ and the bare Coulomb interaction $V$ via a Dyson equation
\begin{equation}
    \label{DysonW}
    W(1,2) = V(1,2) + V(1,\overline{3})
    P(\overline{3},\overline{4})W(\overline{4},2) \;.
\end{equation}
As $\Sigma$, $P$ can be expanded in powers of the screened interaction where to zeroth order in $W$, the RPA is obtained,
\begin{equation}
    \label{rpa}
    P^{RPA}(1,2) = -i G(1,2)G(2,1) \;.
\end{equation}

Solving these Equations fully self-consistently is cumbersome due to the complicated frequency dependence of the interacting Green's function and the second-order self-energy. Since we are interested in a computationally efficient approach, we introduce a few approximations: First, all quantities above are exclusively evaluated with non-interacting Green's functions, which implies that we need to approximate \eqref{DysonG}. Second, $\Sigma^{G3W2}$ is evaluated perturbatively and the static limit for $W$ is used. We will discuss this term in more detail in the next subsection. First, we outline our strategy on the GWA level. 

\subsection{Partial self-consistency in the GWA}

% First, we express all expressions given above in a notation which is more suitable for actual implementations. In the following, summation over repeated indices is implied and we suppress spin-variables throughout. We discretize real space using a set of real atomic orbitals $\mathcal{A} = \left\{\chi_{\mu}(\br)\right\}_{\mu = 1, \dots N_{AO}}$. Molecular orbitals $\mathcal{M} = \left\{\phi_p(\br)\right\}_{p = 1, \dots, N_{MO}}$ are special linear combinations of atomic orbitals, $\phi_p(\br) = b_{\mu p} \chi_{\mu}(\br)$, where $\mu$ runs over all elements of the AO basis, which diagonalize the single-particle Hamiltonian. We will use greek lowercase letters throughout to denote AOs, while roman lowercase letters denote MOs. 

In a discrete basis of canonical molecular orbitals (MO) (we will label general MOs by $p,q,r, \dots$, occupied orbitals by $i,j,k, \dots $, and virtual orbitals by $a,b,c,\dots$), eq. \eqref{DysonG} can be written in the form (see for example ref. \cite{Hirata2017})
\begin{equation}
    \label{DysonH}
    \sum_r \left\{\epsilon_{pr} - \left[ v_{xc} \right]_{pr} + \left[\Sigma_{xc}\right]_{pr}(\omega_p)\right\}U_{rq}(\omega_p) = \omega_{p}U_{pq}(\omega_p) \;.
\end{equation}
Here, $\epsilon$ is the Hamiltonian of a non-interacting reference system, typically calculated within the HF approximation or KS-DFT, and is diagonal in the MO basis. If the off-diagonal elements of $\Sigma_{xc}- v_{xc}$ are neglected, \eqref{DysonH} reduces to a set of independent, non-linear equations,
\begin{equation}
    \label{g0w0}
    \epsilon_p - \left[v_{xc}\right]_{pp} + \left[\Sigma_{xc}\right]_{pp}(\omega_p) = \omega_p \;,
\end{equation}
for the QP corrections $\omega_p$ to the single-electron energy levels $\epsilon_p$. If $\Sigma$ is calculated in the GWA, this approach is referred to as $G_0W_0$. $G_0W_0$ can give very accurate results but this heavily depends on the choice of the underlying exchange-correlation functional.\cite{Bruneval2006, Bruneval2006a, Bruneval2013, Caruso2016, Knight2016} If the eigenvalues of the KS Hamiltonian are too bad of an approximation to the QP energies (for typical GGAs or the LDA, errors are typical of the order of a few eV), the perturbative approach will fail.\cite{Bruneval2006a, Caruso2016} Therefore, it is mandatory to calculate the $G_0W_0$ correction from a potential which is already close to the $GW$ self-energy.\cite{Atalla2013, Hellgren2021} 

In this work, we pursue two closely related and well-known strategies for this: The first one is to perform a $G_0W_0$ calculation using KS eigenvalues and orbitals from a range-separated hybrid calculation as input. This is usually an excellent starting point since the screened Coulomb potential in the RPA reduces to
\begin{equation}
    W(r \rightarrow \infty, \omega = 0) = \frac{e^{-\lambda r}}{r} \;,
\end{equation}
with $\lambda$ proportional to the Fermi-momentum.\cite{Mattuck1992} This is the basic idea behind the construction of range-separated hybrid (RSH) functionals. Solving the KS equations using such a potential emphasizes the importance of screening of the Coulomb interaction for large electron-electron distances and a perturbative correction with eq.~\eqref{g0w0} is usually rather small.\cite{Kanchanakungwankul2021} The range-separation parameter can then also be adjusted in a way that the eigenvalues from the RSH calculation gives very accurate QP energies by itself.\cite{Refaely-Abramson2011, Refaely-Abramson2012, Kronik2012}

The second approach we employ here is QP self-consistent $GW$ (qs$GW$):\cite{Kotani2007, VanSchilfgaarde2006} For a given approximation to the self-energy an optimized KS potential can be obtained by solving the (linearized)\cite{Godby1988} Sham-Schl{\"u}ter equations (SSE)\cite{Sham1983,Casida1995}. With $\Sigma = GW$, the SSE can be solved as in the exchange-only case if the self-energy is approximated as static.\cite{Sharp1953, Talman1976, EngelEberhardandDreizler2013} One then obtains a statically screened exchange (SEX) correction depending on $Re \Sigma(\epsilon^{QP})$. The derivation via the SSE leaves some ambiguity in the choice of the off-diagonal elements. As shown by Ismail-Beigi,\cite{Ismail-Beigi2017} one can also derive an optimized potential via minimization of the gradient of the Klein functional,\cite{Klein1961} leading to the following Hermitian form which has already been suggested by van Schilfgaarde et al.\cite{VanSchilfgaarde2006},
\begin{equation}
    \label{qsgw}
    \left[v^{OP}_{xc}\right]_{pq} = \frac{1}{2} \left\{ 
    \text{Re} \left[\Sigma_{xc}\right]_{pq}(\epsilon_p) + 
    \text{Re} \left[\Sigma_{xc}\right]_{pq}(\epsilon_q)
    \right\} \;.
\end{equation}
Using this potential, the KS equations are then solved self-consistently. It should be stressed that in fully self-consistent GW, an interacting Green's function is constructed in each iteration by solving \eqref{DysonG} while in qs$GW$ a non-interacting Green's function is constructed in each iteration. qs$GW$ should thus rather be seen as an approach to optimize a non-interacting Green's function. It can also be combined with \eqref{g0w0} ($G_0W_0@$qsGW),\cite{Klimes2014a} However, this is typically not done in practice: In situations where the single QP picture is valid, the self-energy will be varying slowly around the QP position\cite{Kopietz1997} and the effect of the $G_0W_0$ correction will vanish.\cite{Lee2020a} qsGW is also closely related to the RPA optimized effective potential (OEP) method when the self-energy is approximated as static in the QP approximation.\cite{Klimes2014a, Riemelmoser2021} Note, however, that only the self-energy is approximated as static, but not the screened interaction, as in Hedin's Coulomb hole (COH)SEX approximation.\cite{Hedin1965}
%Many other variants to determine a RPA potential have been suggested.\cite{Yu2021} For example, it is also possible to solve the \emph{linearized} SSE with the full $GW$ self-energy\cite{Hellgren2007, Hellgren2012, Hellgren2015, Caruso2013b} but the procedure is computationally expensive and has until recently\cite{Riemelmoser2021} only been applied to atoms\cite{Hellgren2007} and small molecules.\cite{Hellgren2012, Hellgren2015, Nguyen2014, Caruso2013b} 

\subsection{Evaluation of the Self-Energy}
We now proceed with the discussion of the $G3W2$ self-energy. To this end, we first introduce the non-interacting Green's function. Since we exclusively work with non-interacting Green's functions we will suppress the superscript $(0)$ in the following. The time-ordered free propagator $G$ is diagonal in the MO basis and can be expressed in terms of greater $G^{>}$ (particle) and lesser $G^{<}$ (hole) propagators,
\begin{equation}
\label{G0}
   G_p(t_1,t_2) = 
   \Theta(t_1-t_2)G_p^{>}(t_1,t_2) +
   \Theta(t_2-t_1)G_p^{<}(t_1,t_2) \;,
\end{equation}
where 
\begin{eqnarray}
\label{G0>}
    G_p^{>}(t_1,t_2) = & G_p^{>}(t_1-t_2) = -i\overline{f}(\epsilon_p) e^{-i\epsilon_p(t_1-t_2)} \\
\label{G0<}
    G_p^{<}(t_1,t_2) = & G_p^{<}(t_1-t_2) = if(\epsilon_p) e^{-i\epsilon_p(t_1-t_2)} \;. 
\end{eqnarray}
Here, $f$ denotes the zero-temperature Fermi-function and $\overline{f} = 1 - f$, and we have indicated that the Green's function only depends on the time-difference at equilibrium. The single-particle energies in \eqref{G0<} and \eqref{G0>} are understood to be relative to the chemical potential which we place in the middle of the HOMO-LUMO gap.

\begin{figure}[hbt!]
    \centering
    \includegraphics[width=0.5\textwidth]{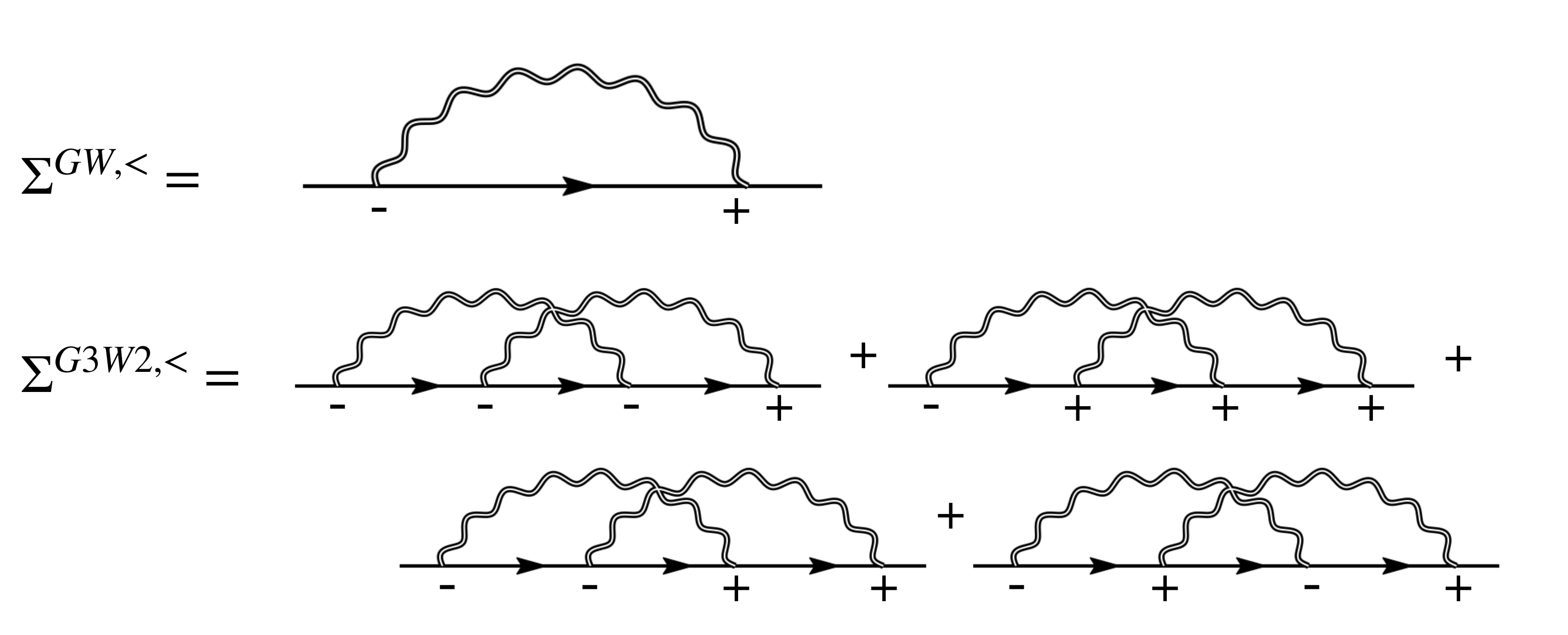}
    \caption{Diagrammatic representation of the lesser components of the $GW$ (top) and $G3W2$ (bottom) self-energy parts. Time-ordering from left to right is implied. The signs at the vertices denote which branches of the Keldysh contour are connected by $W$. Greater components are obtained by exchanging plus and minus signs. Straight lines denote propagators and the double wiggly line is the screened Coulomb interaction.}
    \label{fig::diagrams}
\end{figure}

The structure of the self-energy terms has been scrutinized by Stefanucci, Pavlyukh, van Leeuwen and coworkers.\cite{Stefanucci2014,Pavlyukh2020} We follow their work and use the framework of the Keldysh formalism to discuss the $G3W2$ self-energy.\cite{VanLeeuwen2015} In the Keldysh formalism, we work on the contour $\mathcal{C} = \mathcal{C}^{+} \cup \mathcal{C}^{-}$, with $\mathcal{C}^{+}$ being the backward branch and $\mathcal{C}^{-}$ being the forward branch. The time-ordered (anti-time-ordered) Green's function $G^{--}$ ($G^{++}$) is built from field-operators evolving on $\mathcal{C}^{-}$ ($\mathcal{C}^{+}$) while lesser ($G^{-+} = G^{<}$) and greater $G^{+-} = G^{>}$ Green's function involve both branches and describe propagation of holes and particles, respectively. In the same way, the dynamically screened interaction can either connect both different branches on the Keldysh contour ($W^{+-} = W^{>}$ and $W^{-+} = W^{<}$) or not ($W^{--}$ and $W^{++}$). 

The lesser components of both contributions to the self-energy in \eqref{fullSelfEnergy} are given diagrammatically in figure~\ref{fig::diagrams}. For the $GW$ self-energy each component only consists of a single term (upper part of figure~\ref{fig::diagrams}). For finite systems, the $GW$ self-energy is most conveniently evaluated in a basis of localized atomic orbitals and imaginary time,
\begin{equation}
    \label{Sigma1GW}
    \Sigma^{GW,\lessgtr}_{\mu \nu}(i\tau) =  -i \sum_{\mu' \nu'} G^{\lessgtr}_{\mu'\nu'}(i\tau)W^{\lessgtr}_{\mu\mu'\nu\nu'}(i\tau) \;,
\end{equation}
with $i\tau = i\tau_1-i\tau_2$. In our $GW$ implementation,\cite{Forster2020b} we transform the self-energy to the imaginary frequency axis and analytically continue (either the diagonal elements in the MO basis in $G_0W_0$ or the full self-energy matrix in qsGW) to the complex plane.
Greater and lesser component of the second-order contribution consist of four terms each, since both intermediate vertices can connect both branches of the Keldysh contour. Together, the four diagrams describe three distinct scattering processes.\cite{Pavlyukh2016, Pavlyukh2020} Among others, they are responsible for spectral features which do not appear in fully self-consistent $GW$, like the excitation of two plasmons and two particle hole pairs. In this work, these terms should be of minor relevance only since we are interested in improving QP energies and correlation energies. The last diagram in fig.~\ref{fig::diagrams}, however, only contributes to the $2h1p$ ($1h2p$) space and describes the exchange of two final particles/holes.\cite{Pavlyukh2020} In the same way as the static SOX term it ensures the antisymmetry of the 4-point vertex to first order in the electron-electron interaction. 

Static screening comes with two distinct advantages. Obviously, it drastically reduces the computational cost. Second, only the last diagram in the expression in figure~\ref{fig::diagrams} remains. As an important side effect, the loss of the positive definiteness of the spectral function \cite{Stefanucci2014} arising from straightforward inclusion of all four terms is avoided. The first three diagrams vanish, since a static interaction can only connect points on the same branch of the Keldysh contour.\cite{VanLeeuwen2015} To see this clearly we first write down $\Sigma^{G3W2}$ explicitly,
\begin{equation}
\begin{aligned}
    \Sigma^{G3W2}_{\mu \nu}(\tau_{12}) = & - \sum_{\substack{\kappa\kappa'\lambda\lambda' \\ \mu'\nu'}}\int d \tau_3 d \tau_4
    G_{\mu'\lambda}(\tau_{14})
    W_{\mu\mu'\kappa\kappa'}(\tau_{13}) \\
    & \times 
    G_{\lambda'\kappa}(\tau_{43})
    G_{\kappa'\nu'}(\tau_{32})
    W_{\lambda\lambda'\nu\nu'}(\tau_{42}) \;.
\end{aligned}
\end{equation}
Using a time-independent interaction, the integrals over the internal times can be evaluated easily and transformation to frequency space using
\begin{equation}
    \label{TtoW}
    F(i\omega) = -i \int d\tau F(i\tau) e^{-i\omega \tau}
\end{equation}
gives
\begin{equation}
\begin{aligned}
    \Sigma^{G3W2}_{\mu \nu}(i\omega) = & i \sum_{\substack{\kappa\kappa'\lambda  \lambda'\\\mu'\nu'}} \int d \tau_{12} e^{-i\omega\tau_{12}}
    G_{\mu'\lambda}(\tau_{12})
    W_{\mu\mu'\kappa\kappa'} \\
    & \times 
    G_{\lambda'\kappa}(\tau_{21})
    G_{\kappa'\nu'}(\tau_{12})
    W_{\lambda\lambda'\nu\nu'}\;,
\end{aligned}
\end{equation}
which corresponds to the last diagram in figure~\ref{fig::diagrams} and can be expressed in terms of greater and lesser components of $G$ only (plus and minus signs alternate). Using \eqref{G0}, the integral has a simple analytic solution and transforming to the MO basis gives
\begin{equation}
\label{Sigmag3w2Full}
\begin{aligned}
\left[\Sigma^{G3W2}\right]_{pq}(\omega) = &
\sum_{iab} 
\frac{W_{iapb}W_{ibqa}}
{\epsilon_a + \epsilon_b  - \epsilon_i - \omega}
  \\&  - \sum_{ija} 
\frac{W_{iajp}W_{iqja}}
{\epsilon_a - \epsilon_i - \epsilon_j + \omega } \;,
\end{aligned}
\end{equation}
which is the same expression already employed by Gr{\"u}neis et al.\cite{Gruneis2014} The computational effort to evaluate these terms only grows as $N^3$ with systems size if one is only interested in a small number of diagonal elements. Calculation of the matrix elements of $W$ will generally scale as $N^4$, but these will be available in some form when a $GW$ calculation is performed. The asymptotic scaling can be further reduced by exploiting sparsity in the AO basis or in real space,\cite{Willow2012, Willow2013a,Mardirossian2018} but we do not explore these possibilities here. 

For the self-energy beyond $GW$, we again rely on the assumption that $GW$ already gives rather accurate QP energies. We thus expand $\Sigma^{G3W2}$ around the $GW$ QP energies. At zeroth order, we obtain
\begin{equation}
\label{g3w2_qs}
    \epsilon^{GW + G3W2}_p = 
    \epsilon^{GW}_p + \Sigma^{G3W2}(\epsilon^{GW}_{pp}) \;.  
\end{equation}
This is the most economical way to calculate the QP energy correction due to $\Sigma^{G3W2}$. Note, that on the GWA level, the self-energy is often treated to first order, which results in multiplication with a renormalization factor $Z$. This is commonly done for solids, but it has been argued that the zeroth-order treatment is to be preferred on the GWA level.\cite{Niquet2004} Of course, evaluating \eqref{g0w0} directly with the $\Sigma^{GW + G3W2}$ self-energy is possible as well as has been done in ref.~\citen{Wang2021}. Due to its much smaller computational cost, all values presented in this work have been obtained using equation \eqref{g3w2_qs}.

\subsection{Correlation Energies}
For a given self-energy, total energies can be evaluated using the Klein functional.\cite{Klein1961} Following Dahlen et al.\cite{Dahlen2006} we obtain for the $\Sigma^{GW + G3W2}$ self-energy with a static $W$ in the second-order term  (Trace implies integration over spatial coordinates, spin and frequency variables)
\begin{equation}
\label{corr}
\begin{aligned}
    E_{xc} = & E_x[\phi^{KS}] + \frac{1}{2}\text{Tr}\left\{\ln(1-P^{RPA} V) + P^{RPA} V\right\} \\ & +  \frac{1}{4}\text{Tr}\left\{G_0 \Sigma^{G3W2}_c\right\} \;.
    \end{aligned}
\end{equation}
The first term is the HF exchange-correlation energy expression, evaluated with KS orbitals, the second term is the usual expression for the RPA correlation energy\cite{Langreth1977} and the third term is equivalent to the SOX term in second-order Møller-Plesset perturbation theory (MP2) (see for example eq. (A5) in ref.~\citen{Dahlen2006}), with the bare interaction lines replaces by the statically screened interaction lines. When discussing correlation energies in the following, we refer to the last term on the \emph{r.h.s.} of \eqref{corr} as second-order statically screened Exchange (SOSSX). Correlation energies evaluated with \eqref{corr} is then called RPA+SOSSX, in analogy to RPA+SOSEX.

\section{Technical and Comptational Details}
All expressions presented herein have been implemented in a locally modified development version of the Amsterdam density functional (ADF) engine of the Amsterdam modelling suite 2021.1 (AMS2021). The implementation of the GWA has been outlined in previous work.\cite{Forster2020b, Forster2021, Forster2021a} In all calculations, we expand 2-point correlation functions in correlation consistent bases of Slater-type orbitals of triple- and quadruple-$\zeta$ quality (TZ3P and QZ6P, respectively).\cite{Forster2021} Imaginary time and imaginary frequency variables are discretized using non-uniform bases $\mathcal{T} = \left\{\tau_{\alpha}\right\}_{\alpha = 1, \dots N_{\tau}}$ and $\mathcal{W} = \left\{\omega_{\alpha}\right\}_{\alpha = 1, \dots N_{\omega}}$ of sizes $N_{\tau}$ and $N_{\omega}$, respectively, tailored to each system. More precisely, \eqref{TtoW} is implemented as
\begin{eqnarray}
    \overline{F}(i\omega_{\alpha}) = &  \Omega^{(c)}_{\alpha\beta} \overline{F}(i\tau_\beta) \\
    \underline{F}(i\omega_{\alpha}) = &  \Omega^{(s)}_{\alpha\beta} \underline{F}(i\tau_\beta) \;, 
\end{eqnarray}
where $\overline{F}$ and $\underline{F}$ denote even and odd parts of $F$, respectively. The transformation from imaginary frequency to imaginary time only requires the (pseudo)inversion of $\Omega^{(c)}$ and $\Omega^{(s)}$, respectively. Our procedure to calculate $\Omega^{(c)}$ and $\Omega^{(s)}$  as well as $\mathcal{T}$ and $\mathcal{W}$ follows Kresse and coworkers.\cite{Kaltak2014,Kaltak2014a,Liu2016} The technical specifications of our implementation have been described in the appendix of ref.~\citen{Forster2021}. All 4-point correlation functions are expressed in auxiliary basis sets of Slater type functions which are usually 3 to 5 times larger than the primary bases.\cite{Forster2020,Forster2020b} The transformation between both bases is implemented using the pair-atomic density fitting (PADF) method.\cite{Baerends1973} For an outline of the implementation of this method, we refer to ref.~\citen{Forster2020}. 

We have implemented RPA and RPA+SOSSX correlation energies using \eqref{corr}. In analogy to \eqref{Sigmag3w2Full} and in the same way as in typical MP2 implementations in quantum chemistry codes,\cite{Cooper2017, Forster2020} the last term on the \emph{r.h.s} of 
\eqref{corr} is evaluated directly in the MO basis. The second term on the \emph{r.h.s} of \eqref{corr} is evaluated in the space of auxiliary basis functions, following our $GW$ implementation described in ref.~\citen{Forster2020b}.

\subsection{Energies}
The correlation energies can be calculated with different orbitals. In this work, we use orbitals from HF, PBE0\cite{Ernzerhof1999, Adamo1999}, and the exact exchange (EXX) only OEP\cite{Talman1976, EngelEberhardandDreizler2013} in the implementation of Scuseria and coworkers\cite{Izmaylov2007} within the Krieger Lee Iafrate (KLI)\cite{Krieger1990} approximation.  
\subsubsection{Relative energies}
For relative energies, we use imaginary frequency and time grids of 20 points. Note, that 20 grid points are typically sufficient to converge total energies to within $1e-6$ Hartree.\cite{Kaltak2014} Correlation energies are calculated using,\cite{Helgaker1997}
\begin{equation}
    \label{helgaker}
    E_{CBS} = E_{QZ} - E^c_{QZ} + \frac{E^c_{QZ} * 4^3 - E^c_{TZ} * 3^3}{4^3-3^3} \;, 
\end{equation}
where $E^c_{QZ}$ ($E^c_{TZ}$) denotes the correlation energy at the QZ6P (TZ3P) level and $E_{QZ}$ is the total energy at the QZ6P level. The extrapolation scheme has been shown to be suitable for correlation consistent basis sets but can not be used for KS or HF contributions.\cite{Helgaker1997, Jensen2013} For the purpose of calculating relative energies, we assume these contributions to be converged at the QZ6P level. We use the \emph{VeryGood} numerical quality for integrals over real space and distance cut-offs and the corresponding auxiliary basis set.\cite{Forster2020} Dependency thresholds\cite{Forster2020b} have been set to $5e^{-4}$.

\subsubsection{Total Correlation Energies}
We use 32 points for imaginary time and imaginary frequency each. We use here the TZ3P and QZ6P basis sets with additional diffuse functions as described in ref.~\citen{Forster2021} and additional tight 1s and 2p functions as will be described below.\bibnote{See Supplemental Material at [URL will be inserted by
publisher] for all newly composed basis sets used in this work.} Eq~\eqref{helgaker} is used for extrapolation and numerical quality and auxiliary fit set quality are set to \emph{excellent}. No dependency thresholds have been set. 

\subsection{Quasiparticle Energies}
Charged excitations in this work are calculated using \eqref{g3w2_qs}. For qsGW, we set the dependency threshold to $5e^{-3}$ and perform a maximum of 15 iterations of the self-consistency cycle. Following the recommendations given in ref.~\citen{Forster2021a}, we use \emph{VeryGood} numerical quality and the corresponding auxiliary basis set. We perform $G_0W_0$ calculations using PBE, PBE0, LRC-$\omega$PBEh\cite{Rohrdanz2009} and $\omega$B97-X\cite{Chai2008} orbitals and eigenvalues. We use \emph{Good} numerical quality and the corresponding auxiliary basis set for all $G_0W_0$ calculations and set the dependency threshold to $5e^{-4}$. In ref.~\citen{Forster2020b} these settings have been shown to be appropriate. All QP energies are calculated using
\begin{equation}
    \label{cbsExtra}
    \epsilon_n^{CBS} = \epsilon_n^{QZ} - 
    \frac{1}{N^{QZ}_{bas} }\frac{\epsilon_n^{QZ} - \epsilon_n^{TZ}}{\frac{1}{N^{QZ}_{bas}} - \frac{1}{N^{TZ}_{bas}}} \;,
\end{equation}
where $\epsilon_n^{QZ}$ ($\epsilon_n^{TZ}$) denotes the value of the QP energy using QZ6P (TZ3P) and $N^{QZ}_{bas}$ and $N^{TZ}_{bas}$ denote the respective numbers of basis functions (in spherical harmonics so that there are 5 $d$ and 7 $f$ functions). This expression is commonly used for the extrapolation of $GW$ QP energies to the complete basis set limit for localized basis functions\cite{VanSetten2015}. In ref.~\citen{Forster2021} we demonstrated that using this extrapolation scheme with Slater type basis sets good agreement with other codes is obtained for $GW$ QP energies.

\section{\label{sec::res}Results}

\subsection{Total Correlation Energies}

\begin{figure}
    \centering
    \includegraphics[width=0.5\textwidth]{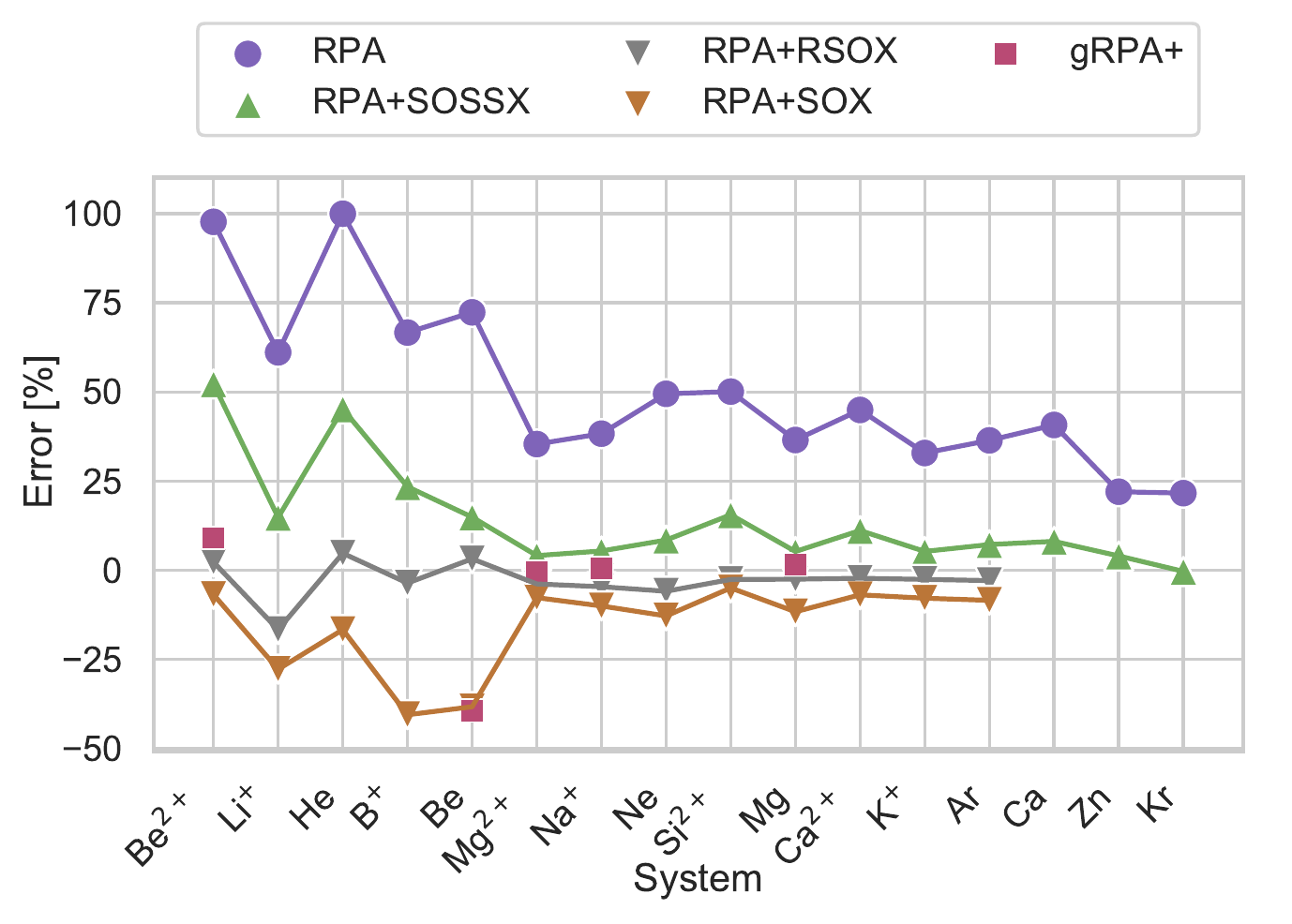}
    \caption{Relative errors [in percent] of correlation energies calculates with different methods compared to the exact values.\cite{Chakravorty1993} The RPA+SOX and RPA+RSOX values are taken from Engel and coworkers\cite{Jiang2007} and the gRPA+ values are from Gould et al.\cite{Gould2019}}
    \label{fig::atoms}
\end{figure}

To see how the inclusion of the $\Sigma^{G3W2}$ self-energy influences the description of electron correlation effects, we first calculate the total correlation energies of 16 atoms with in between 2 and 36 electrons. For all systems with less or equal than 18 electrons, we compare the RPA and RPA+SOSSX correlation energies to almost exact values by Froese-Fischer and coworkers\cite{Chakravorty1993} and for the heavier elements we use the CCSD(T) values by McCarthy and Thakkar as reference.\cite{McCarthy2011} For Argon, their CCSD(T) energy deviates from the value from ref.~\citen{Chakravorty1993} by only 0.01 \%. We also compare them to different beyond-RPA approaches by Jiang and Engel\cite{Jiang2007} (RPA+RSOX and RPA+SOX) and Gould et al.\cite{Gould2019} (gRPA+). To be consistent with ref.~\citen{Jiang2007} and ref.~\citen{Gould2019}, we evaluate the correlation energies with EXX only OEP orbitals (EXX for short), implemented within the KLI approximation.\cite{Krieger1990, Izmaylov2007}

To obtain an idea about the numerical quality of our RPA correlation energies, we compare them against the ones from Engel and coworkers which are free of basis set errors.\cite{Jiang2007} We find deviations between 3 and 15 \% for neutral atoms and much larger ones for cations. Clearly, our standard basis sets are not compact enough for these systems (especially for the cations) and do not capture the full correlation energy. Therefore, we augment them with four tight 1s and two 2p functions each for TZ3P and QZ6P. Using these basis sets, total energies for all atoms deviate to the ones from ref.~\citen{Jiang2007} by about 7 \% on average. This is not perfect but accurate enough for a qualitative comparison of the different beyond-RPA approaches.

The relative errors of correlation energies with respect to the reference values are shown in figure~\ref{fig::atoms}.\bibnote[note1]{See Supplemental Material at [URL will be inserted by
publisher] for all raw data presented in this section.} Simple RPA@EXX overestimates the correlation energies by typically between 25 \% and 100 \%. In accordance with the expectation that the correlation energy is more and more dominated by charge screening with increasing electron number, the agreement with the exact values becomes better for larger atomic numbers. RPA+SOSSX reduced the RPA correlation energy by 29 \% on average,  improving agreement with the reference considerably. Especially for the systems with 2 or 4 electrons, the total correlation energies are still too high. For the last three systems with 18 electrons or more, the agreement becomes much better, but the correlation energy is still overestimated by a small amount. As one can expect, RPA+SOX considerably underestimates the correlation energies. RPA+RSOX shows a tendency to underestimate the correlation energies and out of all assessed methods, the deviations to the reference energies are clearly the smallest. For gRPA+, no clear trend in any direction can be identified.

In summary, the results presented here show that the $GW+G3W2$ self-energy improves the description of electron correlation over $GW$. This, however, does not necessarily imply improvements for properties like relative energies and charged excitations in realistic systems. For example, RPA is often very accurate for relative energies since errors in the correlation energies tend to cancel.\cite{Ren2012a} On the other hand, beyond-RPA methods, while improving total correlation energies, often do not yield improvements for relative energies. For this reason, we shortly assess the performance of RPA+SOSSX in comparison to RPA for relative energies, before we move on to charged excitations.  

\subsection{Energy Differences}

\begin{table}[hbt!]
    \centering
    \begin{tabular}{lccccc}
    \hline 
    & & \multicolumn{2}{c}{RPA} & 
    \multicolumn{2}{c}{RPA+SOSSX} \\
    dataset & $\overline{|\Delta E |}$ & PBE0 &  HF & PBE0 & HF  \\
    \hline 
     ISO  & 21.5 & 1.4 (14.3) & 2.1 (15.1) & 1.0 ( 9.1) & 1.9 (14.1) \\ 
     FH51 & 31.0 & 1.6 (23.6) & 1.6 (21.2) & 1.6 (23.7) & 1.4 (24.0) \\
    \end{tabular}
    \caption{Mean absolute deviations (MAD) and mean absolute percentage deviations (MAPD) (in parantheses) for RPA and RPA+SOSSX using PBE0 and HF orbitals for the ISO and FH51 datasets. The second column shows the average value of the datapoints in each set. All values are in kcal/mol.}
    \label{tab::energies}
\end{table}
We now compare the accuracy of RPA+SOSSX against RPA for relative energies.\bibnotemark[note1] We consider HF and PBE0 starting points. Since we do not evaluate the singles contribution to the total energy, the former is formally more rigorous due to Brillouin's theorem. on the other hand, RPA calculations with PBE0 starting points (RPA@PBE0) are usually more accurate than for HF (RPA@HF). Our benchmark is based on two datasets. The first one consists of 84 relative energies of different conformers of small and medium organic molecules, including transition states of inversion and perycyclic reactions. It is a compilation of the ISO34\cite{Grimme2007b, Goerigk2017}, INV24\cite{Goerigk2016, Goerigk2017}, and BHPERI\cite{Goerigk2010,Goerigk2017, Dinadayalane2002, Guner2003, Ess2005} datasets, with reference values obtained using using the W1-F12\cite{Hylleraas1929, Klopper1987, Kutzelnigg1991a} and W2-F12 protocols\cite{Martin1999, Parthiban2001, Karton2012b}, except for the largest systems in INV24, where reference values have been calculated on the DLPNO-CCSD(T)/\emph{tight}\cite{Riplinger2013, Riplinger2013a, Liakos2015}/CBS level of theory. The second one consists of 51 reaction energies of small organic and inorganic molecules\cite{Goerigk2017, Friedrich2013, Friedrich2015} and the reference values have been obtained on the CCSD(T)-F12/CBS limit of theory.

For RPA@HF and RPA+SOSSX@HF, we report MADs and mean percentage deviations in table~\ref{tab::energies}. The results are promising. PBE0 reference orbitals generally leads to better agreement with experiment, for both, RPA and RPA+SOSSX. Also RPA+SOSSX@HF does not lead to improvements over RPA@HF but RPA+SOSSX@PBE0 leads to significant improvements over RPA@PBE0 for isomerization energies. Relative errors are reduced from 14 \% to 9 \%. For FH51, no improvement is found. 

\subsection{Quasiparticle Energies}
\subsubsection{Organic Acceptor Molecules}
\begin{figure}
    \centering
    \includegraphics[width=0.44\textwidth]{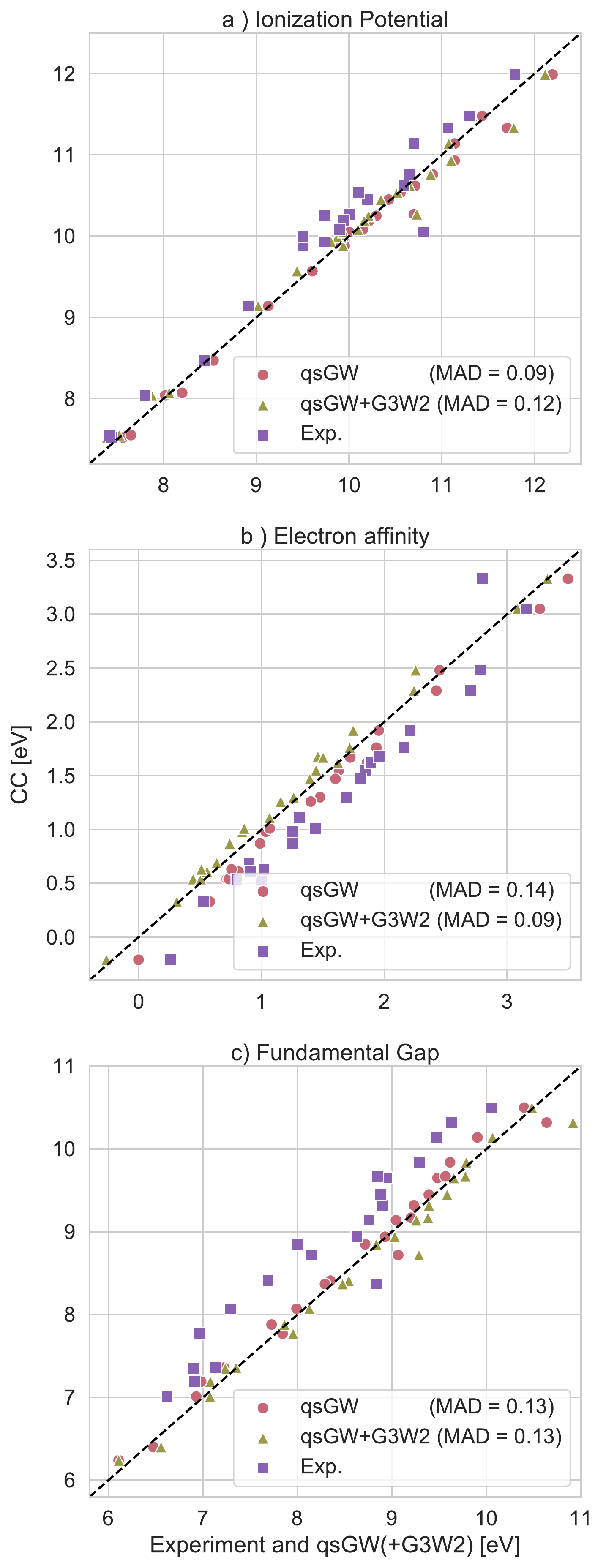}
    \caption{Ionization potentials (top), electron affinities (middle) and fundamental gaps (bottom) from qsGW and qsGW + $G3W2$ for a dataset of 24 organic acceptor molecules. The dashed black diagonal lines are CCSD(T) reference values. Experimental results are given for comparison as well. The values in parantheses denote mean absolute deviations (MAD) with respect to CCSD(T). All values are in eV.}
    \label{fig:acc24_qs}
\end{figure}

We first assess the performance of the $G3W2$ QP energies in comparison to qsGW for a set of 24 organic acceptor molecules.\cite{Richard2016a} The reference values are of CC singles doubles and perturbative triples [CCSD(T)] quality and have been extrapolated to the CBS limit. Comparing these values\bibnotemark[note1] to the experimental data shown in figure~\ref{fig:acc24_qs}, we see that there are sizable differences to the CC values, especially for the fundamental gaps. Among the factors which might contribute to the discrepancies are errors in the optimized geometries, missing zero point vibrational energy corrections, and geometry relaxation after oxidation/reduction. For benzonitrile, the authors of ref.~\citen{Richard2016a} calculated the values of the latter two corrections to be of the order of 0.18 and 0.14 eV, respectively. The errors in geometry or thermodynamical contributions are more difficult to assess but can be sizable as well: For example, the structure used in the calculations might not correspond to a global minimum on the potential energy surface. Finally, we note that the basis set extrapolation can also introduce some errors, especially for the larger systems where no basis sets larger than QZ were used.\cite{Knight2016} We estimate the error of our own CBS limit extrapolation to be of the order of 50 meV for IPs and EAs of medium organic molecules.\cite{Forster2021}. Due to all these factors that affect the direct comparison to experiment, we exclusively use the CCSD(T) reference values for the following quantitative discussion and only show the experimental values for comparison.

\paragraph{Performance of qsGW}

For a variety of solids and metals it has been found that qsGW commonly overestimates band gaps and IPs by about about 15-20 \% when screening is calculated within the RPA.\cite{VanSchilfgaarde2006, Shishkin2007, Kang2010, Svane2010, Punya2011} 
In contrast, the qsGW IPs shown in figure~\ref{fig:acc24_qs}a are in excellent agreement with the CC reference values, with no systematic overestimation. The fundamental gaps in figure~\ref{fig:acc24_qs}c are even (a few exceptions aside) systematically underestimated by qsGW. Overestimation of band gaps with qsGW issue is usually explained by missing electron-hole interaction via vertex corrections in the polarizability. Inclusion of an effective two-point kernel from time-dependent (TD) DFT or the Bethe-Salpeter equation (BSE) has been demonstrated to significantly improve the agreement of band gaps and IPs with experiment,\cite{Shishkin2007, Gruneis2014,  Chen2015, Tal2021, Cunningham2018,Cunningham2021} demonstrating the importance of beyond-RPA screening. For polar materials, i.e. materials with strong longitudinal-optical (LO) and transverse-optical (TO) phonon splitting,\cite{Botti2013} electron-phonon coupling and phonon contributions to the frequency-dependent screening can have a sizable effect on the QP spectrum as well.\cite{Bechstedt2005, Botti2013} For example, qsGW overestimates the experimental band gap of $\text{V}_2\text{O}_5$ by about 100 \%, which to a large extent is due to LO-TO splitting.\cite{Bhandari2015}  

The systems we consider here are rather small and have a planar geometry. The reduction of charge screening in low-dimensional materials has often been emphasized, for example in comparative studies on bulk and layered $\text{V}_2\text{O}_5$\cite{Bhandari2014, Bhandari2015}, $\text{MoS}_2$\cite{Molina-Sanchez2011} or polythiophene.\cite{VanderHorst2000} Anti-screening has been observed in a spin chain\cite{DeGroot1995} and also in finite conjugated systems.\cite{VanDenBrink2000} Finally, LO-TO splitting will be absent entirely. These qualitative differences most likely explain the much higher accuracy of qsGW for the systems studied herein as compared to periodic systems. 

The data clearly demonstrates qsGW to be an excellent first-principle method for the description of charged excitations for these weakly correlated, organic molecules. It is worthwhile to compare the performance of this method for to previous benchmark results of different $GW$ methods. In ref.~\citen{Knight2016}, the accuracy of a large number of $GW$ methods has been assessed for the same dataset. Out of all $GW$ based methods, the authors found $G_0W_0@$LRC-$\omega$PBE to perform best (with optimized range separation parameter), with a MAD of 0.13 eV for IPs, and 0.18 eV for EAs. qsGW seems to be superior.

\paragraph{The effect of the statically screened $G3W2$ correction}

\begin{figure*}[hbt!]
    \centering
    \includegraphics[width=0.92\textwidth]{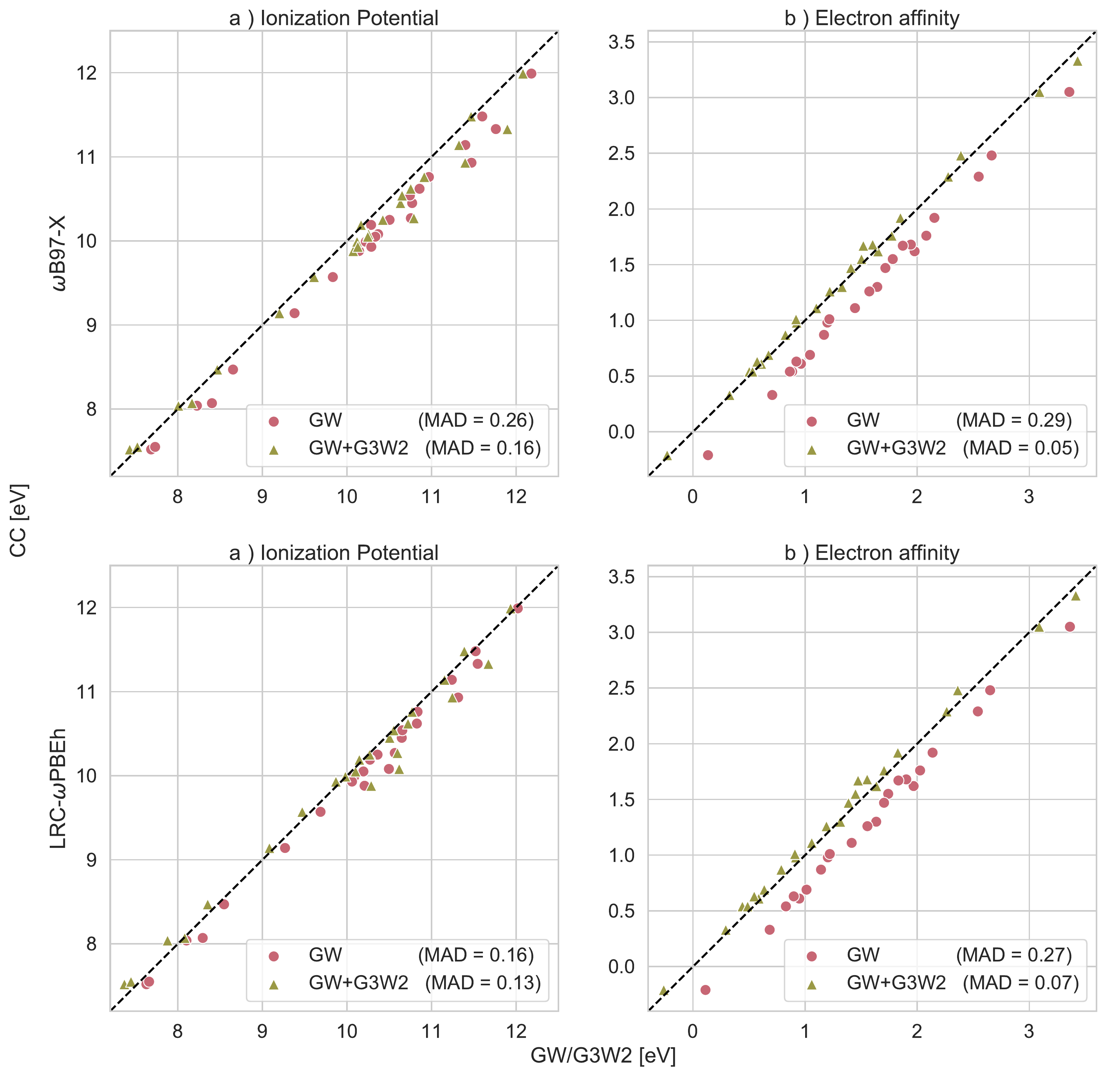}
    \caption{Ionization potentials (left) and electron affinities (right) from $G_0W_0@\omega$B97-X (+ $ \Sigma^{G3W2}$) (top) and $G_0W_0@$LRC$\omega$-PBEh (+ $\Sigma^{G3W2}$) (bottom) and $G_0W_0@\omega$B97-X + $\Sigma^{G3W2}$, for a dataset of 24 organic acceptor molecules. The dashed black diagonal lines are CCSD(T) reference values and the values in parantheses denote mean absolute deviations (MAD) with respect to CCSD(T). All values are in eV.}
    \label{fig:acc24_w}
\end{figure*}

Now we look at the effect of adding the $G3W2$ correction. As can be seen in figure~\ref{fig:acc24_qs}, on average it lowers the qsGW IPs (see also the MDs in table~\ref{tab::MADs}) by a small amount and decreases the EAs by a relatively larger amount, implying increasing fundamental gaps. For IPs, this slightly worsens the agreement with the CC reference values, increasing the mean absolute deviation (MAD) from 0.09 eV to 0.12 eV. However, given the errors from the basis set limit extrapolation, this difference is not significant. Also, the $G3W2$ correction does not alter the MAD for the fundamental gaps. Since the statically screened $G3W2$ correction tends to increase the QP energies, a method which systematically overestimates IPs and EAs will be systematically improved by the $G3W2$ correction. This is demonstrated in figure~\ref{fig:acc24_w} for $G_0W_0@\omega$B97-X and $G_0W_0@$LRC$\omega$PBEh. Here, the $G3W2$ correction slightly improves the MAD with respect to the CC reference values for the IPs from 0.26 to 0.16 eV, and from 0.16 to 0.13 eV. For the EAs, the improvements are tremendous, and the inclusion of the $G3W2$ term leads to almost perfect agreement with the reference values. In table~\ref{tab::MADs}, these results are summarized. Despite the great performance of the $G3W2$ correction for the range-separated hybrids, the description of fundamental gaps is actually deteriorated, which can be considered as a serious drawback of this method.

\begin{table}[]
    \centering
    \begin{tabular}{lcccccc}
    \hline 
    & \multicolumn{3}{c}{{$GW$}} 
    & \multicolumn{3}{c}{{$GW + G3W2$}}   \\
    & qsGW & {$\omega B97X$} & {$\omega$PBEH} & 
    qsGW & {$\omega B97X$} & {$\omega$PBEH}  \\
    \hline
    & \multicolumn{6}{c}{Ionization Potentials} \\
    \hline
MAD      &  0.09 &  0.26 &  0.16 &  0.12 &  0.16 &  0.13 \\
MD       &  0.08 &  0.26 &  0.16 &  0.00 &  0.14 &  0.05 \\
MAX      &  0.43 &  0.54 &  0.41 &  0.46 &  0.57 &  0.54 \\
    \hline
    & \multicolumn{6}{c}{Electron Affinities} \\
    \hline
MAD      &  0.14 &  0.29 &  0.27 &  0.09 &  0.05 &  0.07 \\
MD       &  0.14 &  0.29 &  0.27 & -0.09 & -0.03 & -0.06 \\
MAX      &  0.25 &  0.38 &  0.36 &  0.23 &  0.15 &  0.20 \\
    \hline
    & \multicolumn{6}{c}{Fundamental Gaps} \\
    \hline
MAD      &  0.13 &  0.11 &  0.16 &  0.13 &  0.19 &  0.16 \\
MD       & -0.06 & -0.03 & -0.12 &  0.09 &  0.17 &  0.11 \\
MAX      &  0.35 &  0.25 &  0.28 &  0.60 &  0.66 &  0.62 \\
 \hline
    \end{tabular}
    \caption{Mean absolute deviations (MAD), mean signed deviattions (MD), and maximum errors (MAX) for IPs, EAs, and fundamental gaps for the 24 acceptor molecules for three different starting points for several $GW$ based methods plus the respective $G3W2$ corrections. All values are in eV.}
    \label{tab::MADs}
\end{table}

\begin{table*}[hbt!]
    \centering
    \begin{tabular}{lcccccccc}
    \hline
    Starting Point 
    & \multicolumn{2}{c}{$G_0W_0$}
    & \multicolumn{2}{c}{$G_0W_0$+SOSEX}
    & \multicolumn{2}{c}{$G_0W_0+G3W^{(d)}2$}
    & \multicolumn{2}{c}{$G_0W_0+G3W^{(s)}2$} \\
    & IP & EA & IP & EA & IP & EA & IP & EA \\
    \hline
    PBE               &  0.56/0.66   & 0.44/0.60 & 0.33 & 0.08 & 0.28 & 0.06 & 0.67 & 0.09 \\
    PBE0              &  0.19/0.22   & 0.39/0.39 & 0.12 & 0.16 & 0.16 & 0.09 & 0.28 & 0.06 \\
    \hline
    \end{tabular}
    \caption{Comparison of MADs with respect to the CC reference values for different beyond-$GW$ method in comparison to $G_0W_0$ for different starting points. All values are in eV. The subscript $(d)$ denote that the screened interaction is dynamic, whereas $(s)$ denotes a static interaction. For $G_0W_0$, the two numbers denote MAD with ADF/FHI-AIMS.}
    \label{tab::beyondGWvsGW}
\end{table*}

\paragraph{Comparison of dynamically and statically screened SOX}

In table~\ref{tab::beyondGWvsGW} we compare a few beyond-$GW$ approaches for the same dataset. These are $GW$ + SOSEX from ref.~\citen{Ren2015} and $GW$ + dynamically screened $G3W2$ from ref.~\citen{Wang2021} (denoted as $G_0W_0\Gamma^{(1)}_0$ by the authors of ref.~\citen{Wang2021}). These methods have all been implemented perturbatively and only differ in the way the electron-electron interaction in the SOX-term is screened (see for example figure 1 in ref.~\citen{Wang2021}). The results from refs.~\citen{Ren2015} and ref.~\citen{Wang2021} clearly demonstrate that especially methods with bad or only mediocre performance, like $G_0W_0@$PBE or $G_0W_0@$PBE0, profit immensely from vertex corrections with fully dynamical self-energy: $G_0W_0\Gamma^{(1)}_0@$PBE0 is very accurate, with a MAD of 0.16 eV for the IPs and 0.09 eV for EAs. Especially $G_0W_0\Gamma^{(1)}_0@$PBE performs much better than $G_0W_0$PBE. For EAs, the MAD improve from 0.60 eV to 0.06 meV, however, with 0.28 eV for IPs, $G_0W_0\Gamma^{(1)}_0@$PBE is not very accurate. For EAs, with a MAD of 0.06 eV, $G_0W_0@$PBE + SOSEX performs excellent, but it is considerably less accurate for IPs (MAD = 0.33 eV). In summary, $G_0W_0\Gamma^{(1)}_0@$PBE0 is the most accurate of these methods. However, it can no beat qsGW and also not $G_0W_0@\omega$B-97-X + $G3W2$ and $G_0W_0@$LRC$\omega$PBEh + $G3W2$.

For the sake of a direct comparison of dynamically and statically screened $G3W2$ corrections, we also calculated $G_0W_0@$PB + $G3W2$ or $G_0W_0@$PBE0 + $G3W2$ with our implementation. First, we note the reasonable agreement of the results obtained with our implementation and FHI-AIMS on the $GW$ level, which allows for a qualitative comparison. The differences mainly result from the different basis sets used in the calculations.\cite{Ren2015, Wang2021} The MADs in table~\ref{tab::beyondGWvsGW} clearly show, that the statically screened $G3W2$ correction does not give good results for these starting points. For EAs, the performance of the statically screened correction is comparable to the dynamical one. However, the description of IPs is even worsened. Here, the methods with dynamical screening are significantly better.

In ref.~\citen{Wang2021}, it was found that the magnitude of the $G3W2$ correction was much smaller, when the statically screened interaction instead of the dynamically screened one was used. This is in line with our results. For the HOMO QP energies, the correction often changes sign, i.e. the HOMO QP energy is increased when the interaction is statically screened. Also in this case, the correction for the LUMO level is typically much larger than the one for the HOMO level, increasing the fundamental gaps. Due to the small magnitude of the correction, the statically screened $G3W2$ correction works well if QP energies are already well described on the $GW$ level. The dynamically screened $G3W2$ term leads to a correction of larger magnitude and works best for $GW$ methods which severely underestimate IPs and overestimate EAs. This is the case for $G_0W_0@$PBE and (to a smaller extent) also for $G_0W_0@$PBE0\cite{Knight2016} and consequently, the addition of SOSEX or dynamically screened $G3W2$ leads to large improvements.\cite{ Wang2021} On the other hand, $G_0W_0@$HF underestimates the HOMO QP levels and overestimates LUMO QP levels.\cite{Knight2016, Wang2021} Therefore, the addition of dynamically screened SOX deteriorates the results for this starting point.

\subsubsection{Ionization Potentials of Small Molecules}

\begin{table}[hbt!]
    \centering
    \begin{tabular}{lcc}
    \hline 
    Starting Point 
    & $GW$ 
    & $GW + G3W2$ \\
    \hline
    qsGW             & 0.21 & 0.27 \\
    $\omega$B97-X    & 0.20 & 0.27 \\
    LRC-$\omega$PBEh & 0.12 & 0.21 \\
    \hline 
    \end{tabular}
    \caption{MADs with respect to the EOM-CCSDT reference for the first ionization potentials of a set of 40 small molecules. All values are in eV.}
    \label{tab:gw40}
\end{table}

So far, the performance of the $G3W2$ self-energy correction has only been assessed for a very specific type of molecules. We now also consider a second database curated by Bartlett and coworkers.\cite{Ranasinghe2019} They calculated the IPs of 40 small organic and inorganic molecules, using equation of motion (EOM) CCSD (EOM-CCSD)/cc-pVTZ \& cc-pVQZ and EOM-CCSDT/cc-pVTZ. The reference values we use here are obtained as follows: From the EOM-CCSD results obtained with the cc-pVTZ and cc-pVQZ basis sets, we extrapolate the IPs to the CBS limit with the formula by Helgaker et al.\cite{Helgaker1997}, eq. \eqref{helgaker}. Note, that we used eq.\eqref{cbsExtra} to obtain our results. Subsequently, we add the difference between the EOM-CCSD and EOM-CCSDT IPs to the CBS limit extrapolated EOM-CCSD IPs. Thus, the reference values should be close to EOM-CCSDT quality at the CBS limit. The MADs of the considered methods with respect to the EOM-CCSDT reference values are shown in table~\ref{tab:gw40}. The qsGW IPs are with a MAD of 0.21 eV still in reasonable agreement with the reference values, but the agreement is worse than for the acceptor molecules. For all tested $GW$ starting points, the $G3W2$ term worsens the IPs.

\subsection{Timings}
\begin{table}[hbt!]
    \centering
    \begin{tabular}{lccc}
    \hline
    & Azulene & \multicolumn{2}{c}{Borrelidin} \\
    & QZ6P    & TZ3P & QZ6P \\ 
    \hline 
    $N_{atoms}$&  18            & \multicolumn{2}{c}{78} \\ 
    $N_{freq}$ &  18            & \multicolumn{2}{c}{18} \\ 
    $N_{AO}$   &  782           &  1687 &  3184 \\ 
    $N_{aux}$  &  2296          &  9612 & 14257 \\
    $GW$ calc. [Core h] &           7.3   &    17   & 74  \\
    $G3W2^{*}$ correction [Core h] & 0.015  &  0.40 & 1.6 \\ 
    \hline
    \end{tabular}
    \caption{Computational Timings (in core hours) for two selected molecules for different basis sets. $N_{AO}$ and $N_{aux}$ denote the sizes of primary and auxiliary basis, respectively.}
    \label{tab::timings}
    \raggedright
    \footnotesize{$^*$Timing per state}\\
\end{table}

Before concluding this work, we shortly comment on the computational timings shown in table~\ref{tab::timings}. Compared to SOSEX and $G3W2$ with dynamic $W$, our approach comes with the advantage that it is computationally very cheap. For example, the $G3W2$ correction for Azulene at the QZ6P level can be calculated in 2 seconds each for HOMO and LUMO on a 2.2 GHz intel Xeon (E5-2650~v4) node (broadwell architecture) with 24 cores and 128 GB RAM each, while the preceding $G_0W_0$ calculation takes 1100 seconds. We also calculated QP energies of the Borrelidin molecule (structure taken from ref.~\citen{Ma2020}) with 78 atoms, 266 electrons and 3200 AOs. The $G_0W_0$ calculation alone took 74 core hours, while the $G3W2$ correction took 1.6 core hours per state. Especially if only frontier orbitals are of interest as in many applications, the $G3W2$ correction is thus computationally inexpensive. The comparison is slightly flawed, since our low-scaling $GW$ implementation comes with a higher prefactor than canonical implementations. The $GW$ implementation scales quadratically, while the $G3W2$ corrections is naturally implemented with $N^4$ scaling. Consequently, for much larger systems, The $G3W2$ correction will become computationally more demanding. The corresponding expression for the energy scales as $N^5$ (as opposed to our quadratic scaling RPA implementation) but in canonical implementations MP2 energies can be routinely calculated for systems with more than 100 atoms as well.\cite{Cooper2017} 

\section{\label{sec::conclusion}Conclusion}
In the GWA, the electron-electron self-energy in Dyons's equation is expanded in terms of a screened Coulomb interaction and truncated after first order.\cite{Hedin1965} In this work, we have analysed different aspects of the statically screened second-order contribution to the self-energy ($G3W2$) and applied it in a perturbative fashion to calculate correlation energies of atoms, relative energies of chemical reactions, and IPs and EAs of a wide range of molecules. The results we have presented herein shed some light on many interesting aspects of the second-order correction and also raise many questions which we hope to be able to address in future work.

For correlation energies, the second-order correction (SOSSX) reduces the RPA@EXX correlation energies by around 25 \% and brings them into much better agreement with nearly exact reference values,\cite{Chakravorty1993,McCarthy2011} but the correlation energies are still too large. Generally, relative energies are also improved over RPA, but more detailed investigations are needed to obtain a better understanding in what situations RPA+SOSSX yields improvements over the RPA. 

It would also be interesting to investigate the relation between the RSOX and SOSSX terms further. SOSSX and RSOX are two different strategies to renormalizes the SOX term. RPA+RSOX is based on the resummation of the Epstein-Nesbet series\cite{Szabo2012} of hole-hole ladder diagrams\cite{Engel2006, Jiang2006} while SOSSX results from the resummation of ring diagrams, with the additional approximation of static screening. Both classes of diagrams are contained in CCD which also includes particle-particle ladder diagrams and a third class of diagrams, coupling ladder and ring diagrams.\cite{Scuseria2013} Thus, CCD can be seen as a first-principle method to couple both renormalization strategies. Unfortunately, CCD scales as $N^{6}$, as opposed to $N^{5}$. Since RPA+SOSSX overestimates, and RSOX underestimates the correlation energies, it seems worthwhile to look for alternative ways to combine SOSSX with RSOX while retaining the $N^5$ scaling. 

For charged excitations, our work reveals qualitative differences between the dynamically screened $G3W2$ correction as recently tested in ref.~\citen{Wang2021} and the statically screened one, benchmarked herein. The magnitude of the $G3W2$ correction becomes much smaller when the electron-electron interaction is statically screened. The addition of the dynamically screened SOX term consistently lowers EAs and increases IPs, while the statically screened one often decreases the IPs.

Popular methods like $G_0W_0$ based on PBE and PBE0 starting points predict too low ionization potentials and too high electron affinities\cite{Wang2021} and the addition of dynamically screened $G3W2$ correction results in major improvements for these methods.\cite{Wang2021} The statically screened $G3W2$ correction, on the other hand, gives improvements over $GW$ calculations which consistently underestimate QP energies. This is the case for $G_0W_0@LRC\omega$PBEh and $G_0W_0@\omega$B97-X. Especially for EAs, improvements are tremendous. For IPs and fundamental gaps, the improvements are not consistent and seem to be system-specific. For molecules with nearly 100 atoms, evaluating the perturbative $G3W2$ correction for a few states does not come with significantly increased computational costs compared to a $GW$ calculation. This is a big advantage over the dynamically screened SOX correction.\cite{Wang2021}

Also for qsGW, the statically screened $G3W2$ correction does not lead to systematic improvements, even though both corrections improve fundamental gaps. As an important byproduct of this work, we have shown that qsGW by itself is an excellent method to calculate IPs and EAs for a set of medium organic acceptor molecules. With MADs of 0.09 eV for IPs and 0.14 eV for EAs, qsGW outperforms all other $GW$ methods, previously benchmarked for this dataset.\cite{Knight2016} Remarkably, fundamental gaps are underestimated by qsGW compared to CCSD(T) reference values. This is different from the common situation in extended systems and can be attributed to the much weaker charge screening in finite systems. Also for a second set of 40 small molecules where no reference data for EAs was available, qsGW gives excellent IPs. Charge screening can be expected to be stronger in larger molecules and it would be interesting to see whether qsGW also performs well for these or if beyond-RPA screening is then necessary.

Even though the SOX correction to the self-energy leads to improvements over $GW$ in certain situations, our results confirm recent studies\cite{Lewis2019, Bruneval2021a} which show that QP approximations to $GW$ are difficult to improve upon diagrammatically, at least for molecular systems. Systematic and reliable improvements over $GW$ are most likely only possible starting from the fully self-consistent solution of the $GW$ equations.

\begin{acknowledgments}
This research received funding from the Netherlands Organisation for Scientific Research (NWO) in the framework of the Innovation Fund for Chemistry and from the Ministry of Economic Affairs in the framework of the \emph{TKI/PPS-Toeslagregeling}. 
\end{acknowledgments}

\bibliography{all.bib}

\end{document}